\documentclass[twocolumn,showpacs,prl,floatfix]{revtex4-1}
\usepackage{graphicx}
\usepackage{amsmath}
\usepackage{amssymb}
\usepackage{latexsym}
\usepackage{amsbsy}
\usepackage{array}
\usepackage{amsfonts}
\usepackage{mathrsfs}
\usepackage{color}
\usepackage{hyperref}
\usepackage{mathtools}
\usepackage{footnote}
\usepackage{appendix}

\begin{document}
\title{Geometric chained inequalities for higher- dimensional systems}
\author{Marek \.Zukowski, Arijit Dutta}
\affiliation{Institute of Theoretical Physics and Astrophysics, University of Gda\'{n}sk, 80-952 Gda\'{n}sk, Poland}

\received{\today}

\begin{abstract}
For  systems of an arbitrary dimension,  a  theory of  geometric chained Bell inequalities is presented. The approach is based on chained  inequalities derived by Pykacz and Santos.   For maximally entangled states the inequalities lead to a complete $0=1$ contradiction with quantum predictions. Local realism suggests that the probability for the two observes to have identical results is $1$ (that is,  a perfect correlation is predicted), whereas quantum formalism gives an opposite prediction: the local results always differ. This is so for any dimension. We also show that with the inequalities,  one can have a version of Bell's theorem which involves only correlations  arbitrarily close to perfect ones.  
\end{abstract}

\pacs{}
\maketitle

\newcommand{\bra}[1]{\langle #1\vert} 
\newcommand{\ket}[1]{\vert #1\rangle} 
\newcommand{\abs}[1]{\vert#1\vert} 
\newcommand{\avg}[1]{\langle#1\rangle}
\newcommand{\braket}[2]{\langle{#1}|{#2}\rangle}
\newcommand{\commute}[2]{\left[{#1},{#2}\right]}

\section{introduction}
Bell theorem states that there exists {\em no} local  hidden variable model of quantum theory. The hidden variables are hypothetical additional parameters which are  beyond quantum formalism. They might be interpreted as (local) causes of events or,  more narrowly,  as hidden proper states  of the systems. If one additionally assumes that the probabilities of measurement results depending on such local causes have a Kolmogorovian nature (and that there exists a procedure of random choice of local measurement settings, which is {\em independent} of anything else in the experiment), then  one can derive Bell inequalities of some form.  As individual detection events have, in quantum theory, an inherently  spontaneous nature, such inequalities can be violated by quantum predictions. Also, quantum states do not describe the system but rather our knowledge about their preparation (and subsequent evolution), thus they have entirely different properties than the hypothetical (local) hidden proper states. There is no reason for quantum probabilities to satisfy inequalities based on the hypothesis of existence of the latter ones. If one insists on their existence, or existence of causes, one must abandon either locality  or the independence of settings assumption (often provocatively put as {\em free will}).

Since the pioneering work of Bell,  many new derivations of Bell inequalities have appeared in the literature ( see,  e.g.,  the latest review \cite{STEPHANIE}) and many experimental tests have been  performed, especially in the optical domain \cite{RMP}. Very early,  the so-called chained inequalities were found (see \cite{PEARLE};  however, a more detailed analysis of their statistical behavior was first introduced  in Ref. \cite{BRAUNSTEIN}). A different approach to obtain a similar kind of chained inequality was shown in Ref.\cite{Hardy};  later,  the results were used to construct logical Bell inequalities for qubits~\cite{Abramsky}. The first ones were for dichotomic local outputs only, but later on,  generalizations followed \cite{BARRETT}, \cite{OTHERS} including one in the guise of a ladder Hardy type argument \cite{ADAN}. The procedure of chaining rests on a derivation of an initial inequality, and then,  by upper bounding some of the terms in this inequality by an inequality of a similar kind  involving different settings, one can produce a new one. This iteration can be continued arbitrarily long. 

All of this resembles the geometric triangle inequality for distances, which leads to a quadrangle one and,  by iteration,  to a polygon inequality of as many points as one wishes. In the works of Santos \cite{SANTOS} and Pykacz \cite{PYKACZ},  one can find a derivation of chained Bell inequalities based on geometrical concepts related to  Kolmogorovian probabilities. The aim of our work is to extend their results to multidimensional systems  and to show the full power of the geometric approach.

The derivations shown below are for systems of arbitrary dimensions (for different inequalities of this kind,  see \cite{BARRETT}), and there seems to be no obstacle to the  generalization of the results to an infinite dimension.  However such cases will be studied elsewhere. As a  bright squeezed vacuum resembles,  in many respects,  the Einstein-Podolski-Rosen(EPR) state, such states would probably violate generalizations of the inequalities to infinitely dimensional systems. As a matter of fact, a  squeezed vacuum can be shown to violate a chained inequality of a different kind \cite{ROSOLEK}.
 
Chained inequalities are most interesting if we take into account correlations  close to perfect ones. In this context, one can find a specific application of chained inequalities related to the problems of interpreting Franson-type \cite{FRANSON} two-particle interferometry as a Bell experiment;  see \cite{AERTS}. Here, we also shall concentrate on properties of quantum predictions for our chained inequalities, 
for predictions which are close to perfect correlations. For a very high number of chained settings, we approach a kind of Greenberger-Horne-Zeilinger (GHZ)-type contradiction, like in the case of \cite{Hardy}, \cite{BARRETT}. We also show, that one can use such inequalities to give a rigorous formulation of the heuristic approach to Bell's theorem given in \cite{BALLENTINE}.  One can have a Bell theorem involving only correlations that are  infinitesimally close  to a single perfect one.

\section{Derivations}

Within  Kolmogorov theory of probability, one can introduce for a pair of probabilistic events a notion resembling distance (which can be called probabilistic separation). Let A and B be two events. Then their separation $S(A, B)$ is defined as \cite{SANTOS}
\begin{equation}
\label{separation}
S(A, B)=P(A)+P(B)- 2P(A, B),
\end{equation}where P(A, B) is the joint probability of occurring of  both A and B. Obviously, S(A, B)=S(B, A) and $S(A, B)\geq 0$.
Most importantly S(A, B) satisfies a triangle inequality, \begin{equation}S(A, C)\leq S(A, B)+S(B, C).
\end{equation}This can be derived by using the definition, given in $\eqref{separation}$.The triangle inequality reduces to \begin{equation}
\label{probability}
  P(A, B)+P(B, C)\leq  P(B)+P(A, C)
\end{equation}This relation can be easily proved with Venn diagrams, or other methods. Note that if we have the triangle inequality we can build quadrangle and higher ones. It is important to note that the inequality in $\eqref{probability}$ cannot be used in quantum mechanics, if one is interested in events related to measurement of non-compatible observables. Even for two separated observers, if for example  one assumes that A is an event associated with observable $\hat{A}$ for Alice, say getting the eigenvalue $a'$, and C is an event related to obtain measurement result $c'$ of a different observable $\hat{C}$ also by Alice, while $B$  stands for getting $b'$ when Bob measures $\hat{B}$, then we face the problem that P(A,C) is associated with two non-commensurable observables of Alice, and has no quantum mechanical value. Nevertheless, there is no problem in using the inequality in the context of  (stochastic) local hidden variable theories, as in such a case complementarity does not apply.

Nevertheless, a quadrangle inequality, which is naturally implied by the triangle one, does not face this problem. We can denote Alice's events associated with her choice of settings of the local measuring apparatus by $A_i$ and Bob's by $B_j$, where $i, j\in {0,1}$. We get 
\begin{equation}S(A_0, B_1)\leq S(A_0, B_0)+S(A_1, B_0)+S(A_1, B_1). \label{INEQ-4}\end{equation}
This is just the good-old Clauser-Horne (CH) inequality \cite{CH}:
\begin{eqnarray}&P(A_0, B_0)+P(A_1, B_0)+P(A_1, B_1)&\nonumber\\
&-P(A_1)-P(B_0)-P(A_0, B_1)\leq 0.&\end{eqnarray}
 As it is violated by quantum predictions, we see that the notion of Kolmogorovian probability does not apply to quantum observations (Bohr's complementarity at work).

 Note that we can generalize the above separation-inequality (\ref{INEQ-4}). Let us consider $n$ different experiments on each  sides.  Let us give even indices $i$ to Alice's measurement events at specific local settings,  $A_i$, so that $i=2k$; while for Bob's  events  $B_j$ we shall use odd indices, $j=2k+1$. The following implication of the triangle inequality  holds :\begin{eqnarray}
\label{sep}
&S(A_0, B_{2n-1})\leq S(A_0, B_1)+S(A_2, B_1)&\nonumber\\
&+S(A_2, B_3)+\cdots+S(A_{2n-2}, B_{2n-1})&\nonumber\\&=\sum_{|i-j|=1}{S(A_i, B_j)}.&\label{PYKACZ}\end{eqnarray} This inequality also can be easily written in terms of probabilities. 

However, we would try to derive from the above inequality a distance like inequality for distributions of multivalued variables (assuming that the set of eigenvalues for observables of  Alice and Bob is the same one; this can always be done as eigenvalues related to clicks at specific detectors are question of a convention). Denote by  $S(A^x, B^x)$ the Kolmogorovian separation of the following events: Alice while measuring an observable $\hat{A}$ gets an eigenvalue ${a_x} $, and  Bob, measuring an observable $\hat{B}$, gets an eigenvalue ${b_x}$. To make further notation easier, we assume the following convention for our eigenvalue assignment: ${a_x}={b_x}$ for all $ x=1,\dots, d$. With all that, one can write
\begin{eqnarray}&S(A^x, B^x)=P(A^x)-P(A^x, B^x)+P(B^x)-P(A^x, B^x)&\nonumber\\&=P(A^x,\tilde{B}^x)+P(\tilde{A}^x, B^x),&
\end{eqnarray} where $\tilde{B}^x$ denotes the event of $B^x$ not occurring, similarly $\tilde{A}^x$. Now summing this over all possible $d$ outcomes one gets
\begin{eqnarray}
\sum_{x=1}^{d}{(P(A^x,\tilde{B}^x)+P(\tilde{A}^x,B^x))}=2P(A\neq B),
\end{eqnarray}
 where  $P(A\neq B)$ denotes the probability that if Alice measures $\hat{A}$ while Bob $\hat{B}$, they get {\em different} results.
Obviously:
\begin{equation}
P(A\neq B) = \frac{1}{2}\sum_x{S(A^x,B^x)}.
\end{equation} 
By summing up inequalities \eqref{sep} for all pairs ${A}_i^x$ and ${B}_j^x$, over $x=1,...,d$, we see that $P({A}_i\neq{B}_j)$ satisfy a polygon inequality of the following form 
\begin{equation}
\label{chain}
\sum_{|i-j|=1}{P({A_i}\neq{B_j})}\geq P({A_0}\neq{B_{2n-1}}),\end{equation}
with i=0,2,... 2n-2, and $j=1,3,...2n-1$.
An inequality effectively equivalent to the one above was derived using a different method by Colbeck and Renner for {dichotomic} variables, see e.g. \cite{COLBECK}. 

\subsection{Violation}
Let us apply this inequality to entangled qudits. We shall first  show this for a pair of qutrits, and further on present  a calculation  for an arbitrary dimension. The fact that  two maximally entangled two-qubit states violate chained inequalities, which in the $d=2$ case are equivalent to the ones presented, is well known. 

Chained inequalities work well for measurements which are close to  perfect correlations. Therefore let us first assume that the state that Alice and Bob share is maximally entangled 
\begin{equation}
\label{state}
\left|\psi\right\rangle_{AB}=\frac{1}{\sqrt{3}}(\left|00\right\rangle+\left|11\right\rangle+\left|22\right\rangle) 
\end{equation} 
and the observable that Alice measures is $\hat{A}_0$ has eigenstates 
$\left|0\right\rangle,\left|1\right\rangle$and $\left|2\right\rangle$. 
However Bob's observable $\hat{B}_1$ is $"slightly"$ detuned, with eigenstates 
$\left|0'\right\rangle,\left|1'\right\rangle$and $\left|2'\right\rangle$.
 They satisfy 
\begin{equation}
\hat{U}_B(\theta)\left|j\right\rangle=\left|j'\right\rangle
\end{equation}
and one has a formal equivalence
\begin{equation}
\hat{B}_1=\hat{U}_B(\theta)\hat{B}_0[\hat{U}_B(\theta)]^{-1},
\end{equation}
where in this formula only $\hat{B}_0$ stands for an operator for Bob's subsystem which has in his computational basis the same representation as $\hat{A}_0$ of Alice.
Thus, in such a case 
\begin{equation}
P(i,j)=\left|_A\left\langle i\right|_B\left\langle j'\right|\psi\right\rangle_{AB}|^2,
\end{equation}
which is equal to 
\begin{equation}
P(i,j|A_0, B_1)=\left|_A\left\langle i\right|_B\left\langle j\right|\hat{U}^{-1}_B(\theta)\left|\psi\right\rangle_{AB}\right|^2.
\end{equation}
This is turn, because of the specific properties of the maximally entangled state (\ref{state})
can be put as:
\begin{equation}
P(i,j|A_0, B_1)=\left|_A\left\langle i\right|_B\left\langle j\right|\hat{U}^{*}_A(\theta)\left|\psi\right\rangle_{AB}\right|^2,
\end{equation}
where $\hat{U}_A(\theta)$ stands for an operator for Alice's subsystem which has in her computational basis the same representation as $\hat{U}_B(\theta)$ of Bob.
This is because 
\begin{equation}
\sum_{j=1}^{d}{\left|j\right\rangle_A\hat{U}^{-1}_B(\theta)\left|j\right\rangle_B}=\sum_{i=1}^{d}{\hat{U}_A^*(\theta)\left|i\right\rangle_A\left|i\right\rangle_B}.
\end{equation}
Of course, this a general relation holding for for any unitary transformation. Note, however that for unitary transformations which  are real (orthogonal) one has
\begin{equation}
\sum_{j=1}^{d}{\left|j\right\rangle_A\hat{U}^{T}_B(\theta)\left|j\right\rangle_B}=\sum_{i=1}^{d}{\hat{U}_A(\theta)\left|i\right\rangle_A\left|i\right\rangle_B}.
\end{equation}
From now on, because of this property, we shall use orthogonal $\hat{U}_B(\theta)$.

The next pair of measurements can be $\hat{A}_2$ and 
$\hat{B}_1$ with the eigenstates of $\hat{A}_2$
 given by, 
$\left|i"\right\rangle= \hat{U}_A(\theta)^2\left|i\right\rangle_{A}.$ 
Thus
\begin{equation}
P(i,j|A_2, B_1)=\left|_A\left\langle i\right|_B\left\langle j\right| [\hat{U}^{T}_A(\theta)]^2\hat{U}^{T }_B(\theta)\left|\psi\right\rangle_{AB}\right|^2.\end{equation}
 One has
\begin{eqnarray}
\label{single}
 &\hat{U}^{T}_A(\theta)^2\hat{U}^{T }_B(\theta)\left|\psi\right\rangle_{AB}=\hat{U}^{T}_A(\theta)\left|\psi\right\rangle_{AB},&
\end{eqnarray}
and
\begin{eqnarray}
&P(i,j|A_2, B_1)=\left|_A\left\langle i\right|_B\left\langle j\right|\hat{U}^{T}_A(\theta)\left|\psi\right\rangle_{AB}\right|^2&\nonumber\\
&=\left|_A\left\langle i\right|_B\left\langle j\right|\hat{U}_B(\theta)\left|\psi\right\rangle_{AB}\right|^2.&
\end{eqnarray}

 Let us now introduce specific transformation for the case of two qutrits: 
\begin{equation}
\label{unitary1} 
\hat{U}(\theta)=
\begin{pmatrix}
x & y & z\\
z & x & y\\
y & z & x
\end{pmatrix},
\end{equation}
where $x=\frac{1}{3}(1+2\cos\theta)$, y$=\frac{1}{3}(1-\cos\theta-\sqrt{3}\sin\theta)$,
 $z=\frac{1}{3}(1-\cos\theta+\sqrt{3}\sin\theta)$
 and assume that 
$\theta=\frac{2\pi}{3(2n-1)}$.

An  important property of 
$\hat{U}(\theta)$ is that, $(\hat{U}(\theta))^k=\hat{U}(k\theta)$.
 This is because $\hat{U}(\theta)$ is an orthogonal matrix, representing a rotation with respect to axis given by vector (1,1,1) by an angle $\theta$. 
Obviously in such a case $\hat{U}(\theta)\hat{U}(\theta')=\hat{U}(\theta+\theta')$. 
 Therefore,
\begin{equation} 
\hat{U}_{B}^{2n-1}(\theta)= \hat{U}_B(\frac{2\pi}{3})=
\begin{pmatrix}
0 & 0 & 1 \\
1 & 0 & 0\\
0 & 1 & 0
\end{pmatrix}.
\end{equation}
It transforms the state from 
$\left|0\right\rangle$ to $\left|2\right\rangle$, $\left|1\right\rangle$ to $\left|0\right\rangle$ 
and 
$\left|2\right\rangle$ to $\left|1\right\rangle$.

Now define
 \begin{eqnarray}
 \hat{A}_{k}=\hat{U}_A(\theta)^{k}\hat{A_0}\hat{U}_A(\theta)^{-(k)},\nonumber\\ 
\hat{B}_{k}=\hat{U}_B(\theta)^{2n-1}\hat{A_0}\hat{U}_B(\theta)^{-(k)}.
\end{eqnarray}
The measurement of  $\hat{A}_{2n-2}\otimes\hat{I}$ and $\hat{I}\otimes\hat{B}_{2n-3}$ is giving the following probabilities,\begin{equation}P(i, j|A_{2n-2}, B_{2n-3})=\left|_A\left\langle i\right|_B\left\langle j\right|\hat{U}_{A}^T(\theta)\left|\psi\right\rangle_{AB}\right|^2,\end{equation}
whereas
\begin{equation}
P(i, j|A_{2n-2}, B_{2n-1})=\left|_A\left\langle i\right|_{B}\left\langle j\right|\hat{U}^T_{B}(\theta)\left|\psi\right\rangle_{AB}\right|^2.
\end{equation}
Note that, since  in both  cases $$P(i\neq j)=1-\frac{1}{3}\sum_i|U(\theta)_{ii}|^2,$$ where  $U(\theta)_{ij}$ stand for matrix elements of $\hat{U}(\theta)_{A/B}$, one has for both formulas  \begin{equation}P(i\neq j) =1-\frac{1}{9}(1+2\cos\theta)^2=\frac{1}{9}(8\sin^2\frac{\theta}{2}+4\sin^2{\theta}).\end{equation}

 Concerning the last pair of observables, $\hat{A}_0$ and $\hat{B}_{2n-1}$,  from above  discussion, it is obvious that
\begin{eqnarray}
\label{last}
&\hat{B}_{2n-1}=[\hat{U}_B(\theta)]^{2n-1}\hat{B}_0[\hat{U}_B(\theta)]^{-(2n-1)}.
\end{eqnarray}
The idea is to obtain perfect correlations for the pair of observables $\hat{A}_0,\hat{B}_{2n-1}$ which are completely opposite to the ones for $\hat{A}_0$ and $\hat{B}_0$.  This is the reason why the total angle of `rotation' on Bob's subsystem for the last measurement, $\hat{B}_{2n-1}$, must be $\frac{2\pi}{3}$. This leads to the optimal value of $\theta$  given by $\theta=\frac{2\pi}{3(2n-1)}$. The probabilities read
\begin{equation}
P(i,j|A_0, B_{2n-1})=\left|_A\left\langle i\right|_B\left\langle j\right|\hat{U}^T_{B}(\frac{2\pi}{3})\left|\psi\right\rangle_{AB}\right|^2.
\end{equation}
However
\begin{equation}
\hat{U}^T_{B}(\frac{2\pi}{3})|\psi\rangle_{AB} =\frac{1}{\sqrt{3}}(|02\rangle+|10\rangle+|21\rangle).
\end{equation}
Therefore 
\begin{equation} \label{1}
P(i\neq j|A_0, B_{2n-1})=1.
\end{equation}

Thus summing over all probabilities on left-hand side of the chained-inequality $\eqref{chain}$ and comparing them with the supposedly lower value of the right hand side, which is by (\ref{1}) equal to one, we get, 
\begin{equation}
\frac{N-1}{9}
(8\sin^2\frac{\theta}{2}+4\sin^2{\theta})\geq 1, \label{dimension}
\end{equation}
where $N$ is equal to $2n$. This inequality cannot hold already for $N=2$, and for all higher values of it. Moreover, the  left-hand side tends to zero when $N$ goes to infinity. This is because the rule $\frac{\sin{x}}{x}\rightarrow 1$, for $x\rightarrow 0$, can be applied in both terms.
With $N\rightarrow\infty$ one has $0\geq1$. In this limit the right hand side of the inequality (\ref{chain}), if local realism holds, as it approaches zero,  implies that for measurements of $\hat{A}_0$ and 
$\hat{B}_{2n-1}$ one should expect a  perfect correlation, that is $P(i=j)=1$.  However, quantum mechanics predicts a perfect correlation satisfying $P(i=j+1)=1,$ (modulo 3). We have a kind of GHZ contradiction in the limit of infinitely many infinitely close settings.

\section{arbitrary dimensions}

 Let us  now extend the above results to an arbitrary dimension $d$. The case of $d=2$ is well known, but it can be recovered from what we put here for $d=4$, which we discuss first.

\subsection{Four dimensional systems}

One can get similar results as for $d=3$  with the use of the following simple unitary (orthogonal) matrix:
\begin{equation}\label{4d}
\hat{U}(\theta_1)=
\begin{pmatrix}
\cos\theta_1 &\sin\theta_1 & 0 & 0\\
-\sin\theta_1 & \cos\theta_1 & 0 & 0\\
0 & 0 &\cos\theta_1 &\sin\theta_1\\
0 & 0 &-\sin\theta_1 & \cos\theta_1
\end{pmatrix},~
\end{equation}
where $\theta_1=\frac{\pi}{2(2n-1)}$. After $(2n-1)$ iterations this gives
\begin{equation}\label{PERMUTE}
\hat{U}(\theta_1)^{2n-1}=\hat{U}(\frac{\pi}{2})=
\begin{pmatrix}
0 & 1 & 0 & 0\\
-1 & 0 & 0 & 0\\
0 & 0 & 0 & 1\\
0 & 0 & -1 & 0
\end{pmatrix}.~
\end{equation}
The operation fully permutes initial computational basis states (although this not a cyclic permutation).
We apply these unitary transformation to four dimensional observables on Alice's and Bob's sides; the formal relations of consecutive measurements are the same as in the case of $d=3$.  However, now
 the probabilities entering to each  term on left-hand side of $\eqref{chain}$ are, because of the form of the unitary operation  (\ref{4d}),  equal to $\sin^2\theta_1$. Because of the permutation given by (\ref{PERMUTE}) the  right-hand side of $\eqref{chain}$ is always 1. So, in the end the quantum mechanical values of $\eqref{chain}$ are given by
\begin{equation}
\label{dimension4}
(N-1)\sin^2\left(\frac{\pi}{2(N-1)}\right)\geq 1,
\end{equation}
where $N$ is equal to $2n$. Again this is a contradiction, and with large $N$ it approaches a $0\geq1$ one, which can be given a GHZ interpretation.

\subsection{Higher dimensions}
Dimension $d=5$ hods the key to all higher ones. The  appropriate unitary matrix can be put as
\begin{equation}\label{MATRIX}
\hat{U}(\theta_1, \theta_2)=
\begin{pmatrix} 
a & b & 0 & 0 & 0\\
-b & a & 0 & 0 & 0\\
0 & 0 & x & y & z\\
0 & 0 & z & x & y\\
0 & 0 & y & z & x
\end{pmatrix},~
\end{equation}
 where
 $a=\cos\theta_1$, $b=\sin\theta_1$,
 $x=\frac{1}{3}(1+2\cos\theta_2)$, 
$y=\frac{1}{3}(1-\cos\theta_2-\sqrt{3}\sin\theta_2)$, 
$z=\frac{1}{3}(1-\cos\theta_2+\sqrt{3}\sin\theta_2)$. 
This matrix follows all properties mentioned previously in case of qutrit-rotation, $d=3$, and the qubit case, $d=2$. 
Namely, if we put  the angles $\theta_1$, $\theta_2$ as $\frac{\pi}{2(2n-1)}$ and $\frac{2\pi}{3(2n-1)}$, respectively, we have 

\begin{equation}
\hat{U}(\theta_1, \theta_2)^{2n-1}=\hat{U}(\frac{\pi}{2}, \frac{2\pi}{3})=
\begin{pmatrix}
0 & 1 & 0 & 0 & 0\\
-1 & 0 & 0 & 0 & 0\\
0 & 0 & 0 & 1 & 0\\
0 & 0 & 0 & 0 & 1\\
0 & 0 & 1 & 0 & 0
\end{pmatrix}.~
\end{equation}
This matrix permutes initial computational basis states for d=5. Each of the probabilities in the left hand side of the inequality $\eqref{chain}$ is equal to $1-\frac{1}{5}(2\cos^2\theta_1+\frac{3}{9}(1+2\cos\theta_2)^2)$.

 After adding up all functions in $\eqref{chain}$, we get
\begin{eqnarray}
\label{dimension5}
&(N-1)-\frac{N-1}{5}(2\cos^2(\frac{\pi}{2(N-1)})&\nonumber\\&+\frac{3}{9}(1+2\cos(\frac{2\pi}{3(N-1)}))^2)\geq 1,&
\end{eqnarray}
where again  $N$ is equal to $2n$. This again leads to a contradiction, which with $N\rightarrow\infty$ can be put as $0\geq1$.

We can generalize this approach to an arbitrary dimension, $d$. We can always express any number $d$, which is greater than one, in terms of $2$ and $3$, i.e.,
one can always write 
\begin{equation}
\label{linear} 
d=m2+s3,
\end{equation}
where $s=\frac{d-2m}{3},$  and $m$ and $s$ must be positive integers.
 In such a case,  we can apply in a generalization of (\ref{MATRIX}) a qubitlike transformation to $m$ pairs of dimensions, and 
a qutritlike one to $s$ triples of dimensions.  Of course for odd-dimensional systems the easiest choice is to  put $m$ in such a way  that $s=1$, whereas  for even dimensions one simply  has $m$  qubitlike transformations. The basic unitary operation that we need can be constructed like (\ref{MATRIX}), but now with $m$ $2\times2$  blocks of qubitlike form (defined by $a$ and $b$) , and the last  $3\times3$ block just as in (\ref{MATRIX}). Obviously,  $2n-1$ applications of such a matrix leads to a complete permutation of basis states.
 Under such operations the  chained inequality (\ref{chain}) leads to
\begin{eqnarray}
\label{dimensionD}
&(N-1)-\frac{N-1}{d}(2m\cos^2(\frac{\pi}{2(N-1)})&\nonumber\\&+\frac{d-2m}{9}(1+2\cos(\frac{2\pi}{3(N-1)}))^2)\geq 1.&
\end{eqnarray}
In $\eqref{dimensionD}$, as $N$ tends to infinity left-hand side of the inequality goes to zero, although right-hand side is always $1$. So, distance-like inequalities for local realistic description are violated by quantum mechanics. As before local realistic prediction based of the left hand side implies a perfect correlation of a completely different kind, $P(i=j|\hat{A_0}, \hat{B}_{2n-1})=1$, than quantum prediction for the right hand side measurements - also a perfect correlation
but with  $P(i=j|\hat{A_0}, \hat{B}_{2n-1})=0$

\section{Generalization}

These results can be further amplified.  We can abandon the constraint to  our walk on the polygon, which is in the case of inequality (\ref{chain}), 
from $A_0$ to $B_1$, next from $B_1$ to $A_2$, and so on until we reach the next-to-last  step from $A_{2n-2}$ to $B_{2n-1}$ (all this a "longer way" than directly from $A_0$ to $B_{2n-1}$). We can add one more step from $B_{2n-1}$ to $A_{2n}$, and compare this with the separation of the first and the last event that is $A_0$ and $A_{2n}$. In this way we get an inequality which holds for local hidden variables in the form of:
\begin{equation}
\label{chainB}
\sum^{i=2n, j=2n-1}_{|i-j|=1}{P({A_i}\neq{B_j})}\geq P({A_0}\neq{A_{2n}}),\end{equation}
with i=0,2,... 2n, and $j=1,3,...2n-1$. At first glance this inequality seems as useless in quantum mechanics as the triangle one. However, if $A_0 $
and $A_{2n}$ are compatible, that is they commute, it can be compared with quantum predictions. The idea therefore is to use transformations $U_1$ which after $2n$ applications, that is for $U_1^{2n} $, give a permutation of the original basis (this would mean for the ones introduced earlier putting $\theta_1$, $\theta_2$ as $\frac{\pi}{2(2n)}$ and $\frac{2\pi}{3(2n)}$). One can repeat all reasonings given earlier to get a $0=1$ contradiction for  $P({A_0}\neq{A_{2n}})$. This implies directly an absolute contradiction in the local hidden variable prediction, as for any theory one must definitely have $P({A_0}\neq{A_{2n}})=1$ since the difference between the two observables is just a permutation of the results (eigenvalues).  Compare \cite{COLBECK}, where such a contradiction is explicitly  shown for only $d=2$. 

However,  we can start all that with an arbitrary $\hat{A}'_0$, redefine the computational basis such that it is now built out of eigenstates of $\hat{A}'_0$, and find a ``conjugate" $\hat{B}'_0$, such that its eigenstates enter the Schmidt decomposition of the maximally entangled state involving eigenstates of   $\hat{A}'_0$ (recall that a maximally entangled state has infinitely many equivalent Schmidt decompositions). With this we can repeat all the reasonings given above. This leads us to an absolute internal contradiction for a hidden variable description of any observable.  

As a matter of fact, one can derive a kind of Zeno paradox for any local  hidden variable description of observables describing a maximally entangled state. With the construction like above, even if   $\hat{A}_0$ and $\hat{A}_{2n}$ are incompatible (in quantum theory), a local hidden variable theory must give a definite prediction for  $P({A_0}\neq{A_{2n}})$. If these are two {\em different} observables, one must have $P({A_0}\neq{A_{2n}})> 0$, because  $P({A_0}={B_{2n}})< 1$ and  $P({A_{2n}}={B_{2n}})= 1.$ However  for a reasoning like above, in the limit of infinitesimally slow changes of the observables into the consecutive ones, in the chained inequality the left hand side always tends to zero, implying $P({A_0}\neq{A_{2n}})=0$. That is up to sets of (probability) measure zero one has identical local hidden variable models of the two observables. We have no change if we move by infinitesimally small steps, even if they accumulate to a finite one.  Thus reasoning involving perfect correlations leads to absolutely absurd contradictions for local hidden variable models.

\section{Contradiction involving neighborhood of one perfect correlation}

Let us consider a maximally entangled state for a pair of qudits:
\begin{equation}
|\psi\rangle=\frac{1}{\sqrt{d}}\sum_{k=0}^{d-1}|kk\rangle
\end{equation}
The measurements that we shall consider will have two traits. First of all, they will be very close to one giving perfect correlations, and the unitary transformations leading us to other  measurement settings would be constrained to just first two basis vectors of each of the the systems. Thus we have in this sector basically SU(2) transformations. The probabilities  $P(\lambda_{A_i}\neq\lambda_{B_j})$ can in such circumstances be put as 
\begin{equation}
 P(\lambda_{A_i}\neq\lambda_{B_j})=\frac{2}{d} P_q(\lambda_{A'_i}\neq\lambda_{B'_j}) 
\end{equation}
where the primed observables are effective qubit observables describing the effects of the measurements, and the probabilities $P_q$  are the ones for a two qubit system
which effectively describes the sector in which work our constrained  transformations.

In further considerations we shall drop the subscript $q$  and the primes. Thus our calculations will be presented like if we are considering 
a two-qubit system in  $\phi^+$ Bell state  state
\begin{equation}
|\psi\rangle=\frac{1}{\sqrt{2}}\sum_{k=0}^{1}|kk\rangle
\end{equation}
We shall use the spin 1/2 approach to qubits, with local measurements described by the Pauli operators $\vec{a}\cdot\vec{\sigma}_1$  and  $\vec{b}\cdot\vec{\sigma}_2$, where $\vec{a}$ and $\vec{b}$ are the local Bloch vectors defining the measurement direction.
In such a case 
the quantum predictions for measurement results of Alice, $\lambda_A=\pm1$ and Bob, $\lambda_B=\pm1$, are given by
\begin{equation}
P(\lambda_A,\lambda_B)=\frac{1}{4}(1+ \lambda_A\lambda_B\vec{a}\cdot \hat{T}\vec{b}).
\end{equation}
 In the case of $\phi^+$ state the  correlation tensor on z-x plane is written as,
\begin{equation}
\label{CHAIN4}
\hat{T}=\vec{z}\otimes \vec{z}+\vec{ x}\otimes\vec{ x} -\vec{ y}\otimes\vec{ y}.
\end{equation}
If we use Bloch vectors defining the local settings, $\vec{a}$  and $\vec{b}$, which are constrained to the z-x plane only the first two terms matter.

A chained-inequality for a  pair of qubits with $2n$  settings reads
\begin{equation}
\label{CHAIN}
\sum_{|i-j|=1}{P(\lambda_{A_i}\neq\lambda_{B_j})}\geq P(\lambda_{A_0}\neq\lambda_{B_{2n-1}}),
\end{equation}
where $\lambda_{A_i},\lambda_{B_j}$ are the measurement outcomes on Alice and Bob's side respectively while measuring observables $A_i$ and $B_j$.

For a pair of observables with  dichotomic outcomes one has
\begin{equation}
 P(\lambda_{A_i}\neq\lambda_{B_j})=\frac{1}{2}(1-\vec{a}.\hat{T}\vec{b}),
\end{equation}
where $\vec{a}, \vec{b}$ are measurement directions for Alice and Bob respectively.  Using the above  $\eqref{CHAIN}$ can be written in the following form
\begin{eqnarray}
 &\hat{T}\cdot((\vec{a_0}+\vec{a_2})\otimes \vec{b_1}+(\vec{a_4}&\nonumber\\&+\vec{a_2})\otimes \vec{b_3}\cdots+(\vec{a}_{2n-2}-\vec{a_0})\otimes \vec{b}_{2n-1})&\nonumber\\ &\leq (2n-2).&
\end{eqnarray}

Assume the following settings
\begin{eqnarray}
\label{CHAIN1}
&\vec{a}_{2i-2}=\vec{z}\cos(\frac{\pi(2i-2)}{\gamma(2n)})+\vec{x}\sin(\frac{\pi(2i-2)}{\gamma(2n)})&\nonumber\\
&\vec{b}_{2i-1}=\vec{z}\cos(\frac{\pi(2i-1)}{\gamma(2n)})+\vec{x}\sin(\frac{\pi(2i-1)}{\gamma(2n)})&
\end{eqnarray}
Now, each pair of consecutive direction vectors are separated by the same angular separation, $ (\frac{\pi}{\gamma}\frac{1}{2n})$ . So, they follow the relation,
\begin{equation}
\label{CHAIN2}
 \vec{a}_{2k}+\vec{a}_{2k+2}=2\vec{b}_{2k+1}\cos(\frac{\pi}{\gamma}\frac{1}{2n}),
\end{equation} where $k\in\{0, 1\cdots , (n-2)\}.$ Next we insert $\eqref{CHAIN2}$ in $\eqref{CHAIN}$ to obtain a compact form,
 \begin{eqnarray}
\label{CHAIN3}
 & 2\hat{T}\cdot(\cos(\frac{\pi}{\gamma}\frac{1}{2n})\sum_{i=1}^{i=n-1}{ \vec{b}_{2i-1}\otimes \vec{b}_{2i-1}})&\nonumber\\&+\hat{T}\cdot(\vec{a_{2n-2}}-\vec{a_0})\otimes \vec{b}_{2n-1}\leq 2n-2,&
\end{eqnarray}
 Now, using $\eqref{CHAIN4}$, the left-hand side of $\eqref{CHAIN3}$ is reduced to
\begin{eqnarray}
\label{CHAIN5}
 &2\cos(\frac{\pi}{\gamma}\frac{1}{2n})(\sum_{i=1}^{i=n-1}\cos^2(\frac{\pi(2i-1)}{\gamma(2n)})&\nonumber\\
&+\sum_{i=1}^{i=n-1}\sin^2(\frac{\pi(2i-1)}{\gamma(2n)}))+\hat{T}\cdot(\vec{a}_{2n-2}-\vec{a_0})\otimes \vec{b}_{2n-1}\leq 2n-2, &\nonumber\\
\end{eqnarray}
and this of course reduces to
\begin{eqnarray}
& 2(n-1)\cos(\frac{\pi}{\gamma}\frac{1}{2n})+\hat{T}\cdot(\vec{a}_{2n-2}-\vec{a_0})\otimes \vec{b}_{2n-1}\leq 2n-2.&\nonumber\\
\end{eqnarray}

Our next task is to estimate the value of $\hat{T}\cdot(\vec{a}_{2n-2}-\vec{a_0})\otimes \vec{b}_{2n-1}$.  According to $\eqref{CHAIN1}$,
\begin{eqnarray}
\label{CHAIN6}
&\vec{a}_0=\vec{z}, &\nonumber\\
&\vec{a}_{2n-2}=\vec{z}\cos(\frac{\pi(2n-2)}{\gamma(2n)})+\vec{x}\sin(\frac{\pi(2n-2)}{\gamma(2n)}), &\nonumber\\
&\vec{b}_{2n-1}=\vec{z}\cos(\frac{\pi(2n-1)}{\gamma(2n)})+\vec{x}\sin(\frac{\pi(2n-1)}{\gamma(2n)}). &
\end{eqnarray}
Thus,
\begin{eqnarray}
\label{CHAIN7}
&\hat{T}\cdot(\vec{a}_{2n-2}-\vec{a_0})\otimes \vec{b}_{2n-1}=(\cos(\frac{\pi(2n-2)}{\gamma(2n)})-1)\cos(\frac{\pi(2n-1)}{\gamma(2n)})&\nonumber\\
&+\sin(\frac{\pi(2n-2)}{\gamma(2n)})\sin(\frac{\pi(2n-1)}{\gamma(2n)})&\nonumber\\
&=\cos(\frac{\pi}{\gamma(2n)})-\cos(\frac{\pi(2n-1)}{\gamma(2n)})&
\end{eqnarray}
After adding up all terms in $\eqref{CHAIN3}$, we get,
\begin{eqnarray}
\label{CHAIN8}
&2(n-1)\cos(\frac{\pi}{\gamma}\frac{1}{2n})&\nonumber\\
&+\cos(\frac{\pi}{\gamma(2n)})-\cos(\frac{\pi(2n-1)}{\gamma(2n)})\leq 2n-2&\nonumber\\&
\end{eqnarray}
or
\begin{eqnarray}
  & 2(n-1)(\cos(\frac{\pi}{\gamma}\frac{1}{2n})-1)\nonumber&\\
&+\cos(\frac{\pi}{\gamma(2n)})-\cos(\frac{\pi(2n-1)}{\gamma(2n)})\leq 0&
\end{eqnarray}
But, when $n$ tends to infinity for any fixed finite $\gamma$ left-hand side of $\eqref{CHAIN8}$ is sooner or later greater than zero. Hence, the inequality in $\eqref{CHAIN}$ is violated without going through the entire Bloch sphere. For very large $\gamma$,  the derivation involves basically only perfect correlations. Thus we have a kind of an approximate GHZ contradiction for maximally entangled two- system states - as it is based on  correlations which can be infinitesimally close to a (single) perfect one. However, it is not  "all or nothing". Nevertheless,  the interesting aspect is that it is based on correlation in the "epsilonic" neighborhood of the one single perfect one. This might be seen as a rigorous version of the heuristic argumentation given in Ballentine's  textbook (Ref.~\cite{BALLENTINE}, p. 587),  and,  as a bonus, one working for system of arbitrary dimensions.

\section{Conclusions}
The main results of our work can be summarized as follows. The Pykacz-Santos chained inequalities can be generalized to situations in which we have entangled systems of arbitrary dimension. This,  in turn,  in the limit of infinitely equally spaced settings, leads to a no-go theorem for a local realistic description involving perfect correlations only. Another result is that one can also have conclusions of a similar kind involving only correlations infinitesimally close to just one in which we have a perfect correlation.

As the discussed inequalities are valid for any dimension, the results can also  be applied to the case of the dimension of the systems approaching infinity.
In a forthcoming work we shall analyze  the so-called bright squeezed vacuum (BSV) with the methods presented here, seeking drastic consequences for the hypothesis of local realism. Note that the  BSV is a (physical) approximation of  the original (unphysical) EPR state.

\section{Acknowledgments} M.Z.  is supported by BRISQ2 EU project, while A.D. is supported  by an MPD Project of FNP.


\begin{thebibliography}{20}
\bibitem{STEPHANIE} N. Brunner, D. Cavalcanti, S. Pironio, V. Scarani, and S. Wehner,  Rev. Mod. Phys. {\bf 86},  419 (2014).
\bibitem{RMP} J.-W. Pan, Z.-B. Chen, C.-Y. Lu, H. Weinfurter, A. Zeilinger, and M. \.Zukowski, Rev. Mod. Phys. {\bf 84},  777 (2012).
\bibitem{PEARLE}
 P. A. Pearle, Phys. Rev. D {\bf 2}, 1418 (1970).
\bibitem{BRAUNSTEIN}
 S. L. Braunstein and C. M. Caves, Ann. Phys. (NY) {\bf 202},
22 (1990).
\bibitem{Hardy}L. Hardy,  Phys.  Lett. A {\bf 161},(1991)21-25.
\bibitem{Abramsky}S. Abramsky, L. Hardy, Phys. Rev. A {\bf 85}, 062114(2012).
\bibitem{BARRETT}J. Barrett, A. Kent, and S. Pironio,  Phys. Rev. Lett.
{\bf 97}, 170409 (2006).
\bibitem{OTHERS} P. Kurzynski, D. Kaszlikowski, Phys. Rev. A {\bf 89}, 012103 (2014).
\bibitem{ADAN} A. Cabello, Phys.Rev. A {\bf 58}, 1687 (1998).
\bibitem{SANTOS} E. Santos, Phys. Lett. A {\bf 115}, 363 (1986).
\bibitem{PYKACZ}
J. Pykacz and E. Santos, J. Math. Phys. {\bf 32} , 1287 (1991).
\bibitem{ROSOLEK} K. Rosolek, M. Stobinska, M. Wiesniak, M. Zukowski (unpublished).
\bibitem{FRANSON} J. Franson,  Phys. Rev. Lett. {\bf 62}, 2205 (1989).
\bibitem{AERTS} S. Aerts, P. Kwiat, J.-A. Larsson, and M. Zukowski,  Phys.
Rev. Lett. {\bf 83}, 2872 (1999).
\bibitem{BALLENTINE} L.E. Ballentine, {\em Quantum Mechanics: A Modern Development} (World Scientific, Singapore, 1998).

\bibitem{CH} J. Clauser and M. Horne,  Phys. Rev. D {\bf 10}, 526 (1974).

\bibitem{COLBECK}
R. Colbeck, R. Renner
Nat.  Commun. {\bf 2}, 411 (2011);
 Phys. Rev. Lett. {\bf 101}, 050403 (2008).

\end{thebibliography}
\end{document}